\documentstyle[12pt,world_sci]{article}
\begin{document}
\begin{flushright}
INFNCA-TH-94-17\\
August 1994
\end{flushright}

\title{{\bf DIMENSIONAL REDUCTION FOR FERMIONS%
\thanks{Talk presented by M.~Lissia at the ``Workshop on Quantum
Infrared Physics'', Paris, France, 6--10 June 1994.
To appear in the Proceedings.}
}}
\author{
           MARCELLO LISSIA\\
            {\em
   Istituto Nazionale di Fisica Nucleare, via Negri 18, Cagliari, \\
            I-09127, ITALY
            }\\
           \vspace{0.3cm}
           and \\
           \vspace{0.3cm}
            SUZHOU HUANG\\
            {\em
   Department of Physics, FM-15, University of Washington, Seattle, \\
   Washington 98195, U.S.A.
            }
        }
\maketitle
\setlength{\baselineskip}{2.6ex}

\begin{center}
\parbox{13.0cm}
{ \begin{center} ABSTRACT \end{center}
  {\small
We generalize the concept of dimensional reduction to operators
involving fermion fields in high temperature field theories. It is
found that the ultraviolet behavior of the running coupling constant
plays a crucial role. The general concept is illustrated explicitly
in the Gross-Neveu model.
  }
}
\end{center}
\section{Introduction}
Physics very often simplifies under certain extreme situations.
For example, specific observables in a $D+1$-dimensional theory
at high temperature ($T$) can sometimes be described by
a $D$-dimensional theory: this phenomenon is known as dimensional
reduction (DR).
The basic concept of DR is that in the high-$T$ limit all
spatial excitations are naturally ${\cal O}(T)$. If, for either
kinematic or dynamical reasons, there exist modes of order less
than $T$, these few light modes may be the only active ones, while
the others decouple.
This qualitative expectation can be formalized
\cite{appelquist} in some theories, such as QED, order by order
in perturbation theory, analogously to the usual heavy mass decoupling.

The approach to DR is easier, when there exists a clear scale separation
already at the tree-level. For instance, kinematics makes this scale
hierarchy manifest for observables made by elementary boson fields.
The non-zero Matsubara frequencies act like masses of ${\cal O}(T)$
compared to the zero-modes, and the heavy non-zero modes, both bosonic and
fermionic, can be integrated out. If no other dynamical phenomena occur,
the result is an effective theory in one dimension less,
which can be used to describe static phenomena of the original theory
in the high-$T$ limit. So far, the existing literature has exclusively
dealt with the dynamics of these bosonic zero modes.

This work will focus on the situations where the observables are made
explicitly by fermion fields, and, therefore, the lowest modes are also
of ${\cal O}(T)$, and there is no obvious kinematic scale separation.
If at all, the scale separation must be generated dynamically.
We shall systematically study, in the frame work of perturbation theory
and renormalization group, how the concept of DR can be generalized
to include these cases. Our result will provide a formal basis for the
recent interpretation of lattice data related to screening processes
at high-$T$ \cite{koch,hansson,schramm}, which assumes a picture
where only the lowest Matsubara frequencies are important and can be
treated non-relativistically.
After stating the precise criterion for DR for
observables involving fermion fields, we illustrate the general principle
by explicitly calculating the screening mass in the Gross-Neveu model.
\vspace*{0.1cm}
\section{Dimensional reduction}
Let us first recapitulate the criterion for DR when only static
fundamental bosons appear in the external lines.
If Green's functions with small external momenta ($|{\bf p}|\ll T$)
in a theory described by the $D+1$ dimensional Lagrangian
${\cal L}_{D+1}$ are equal to the corresponding Green's functions of
some specific $D$ dimensional Lagrangian ${\cal L}_{D}$, up to corrections
of the order $|{\bf p}|/T$ and $m/T$, where $m$ is any external
dimensionful parameter in ${\cal L}_{D+1}$, e.g. a mass, we say that
DR occur for ${\cal L}_{D+1}$ with ${\cal L}_D$ as the effective theory.
In general, the form and parameters of ${\cal L}_{D}$ are determined
by the original theory. As stressed by Landsman\cite{landsman},
this expectation may fail if there are dynamically generated scales
of the order $T$. Nevertheless, these dynamically generated scales must
be proportional to some power of the coupling constant $M\sim g^n T$,
since they are generated by the interaction. Therefore, they
induce corrections of order $M/T\sim g^n$. If $g$ is small, the concept
is still useful, and we say that the reduction is partial.

The case when composite operators made by fermions appear in the
external lines is of great phenomenological interest, and includes
external currents made out of fundamental fermionic fields such
as the electromagnetic current, or mesonic interpolating fields in QCD.
When fermions appear in external lines there are two main differences
with the fundamental bosons case. The first difference is that the
lowest Matsubara frequencies for fermions, $\omega_\pm=\pm\pi T$,
are also the order $T$, and hence it is not obvious that they dominate
over the heavier modes. The second difference is that we need to consider
external momenta of order $T$. In fact, if we want the fermions to be
close to their mass shell in the reduced theory (this is eventually the
physically relevant region), $|{\bf p}|$ must be the order $T$.

Since $\omega_\pm$ acts in the reduced theory as a large mass,
it has been proposed\cite{koch,hansson,schramm} that fermions might
undergo a non-relativistic kind of dimensional reduction. This motivates
us to define $q^2\equiv {\bf p}^2 -(\pi T)^2$, and expand the
Green's functions in the dynamical residual momentum $q^2$.
In the end, it will be necessary to check the consistency of this
expansion by verifying whether $q^2 \ll (\pi T)^2$. Again,
we expect corrections of order $m/T$ and $q/T$, and talk of
partial reduction if we find that $m$ and/or $q$ is
proportional to $g^n T$, with $g$ small.

In analogy with the heavy mass decoupling theorem, the decoupling of the
heavier modes is manifest only in specific subtraction schemes, such as
the BPHZ scheme. A two-step approach better illustrates the need for
a judicious choice of the counterterms.

Let us consider a graph in the original theory renormalized
in a $T$-independent scheme, e.g. the $\overline{\rm MS}$ scheme.
We can always split it into light and heavy contributions:
$G^{D+1}(q,T) = G_{L}^{D}(q,T) + G_{H}^{D+1}(q,T)$, where $ G_{L}^{D}$
is the contribution of terms where {\em all\/} loop frequencies have
their smallest value. Since there is no infinite frequency sums
$G^D_L$ is actually $D$ dimensional. Then, we expand $G_{H}$ at $q=0$
(the leading infrared behavior is contained in $G_{L}$ by construction),
and keep terms that are not suppressed by powers of $T$:
$G^{D+1}(q,T) = G_{L}^{D}(q,T) + G_{H}^{D+1}(0,T) + O(q/T)$, where we
have supposed that only the first term in the expansion survives. When
DR takes place, the local term $G_{H}^{D+1}(0,T)$ can be eliminated
by changing the renormalization prescription (i.e. adding finite
counterterms), and we are left with the reduced graph $G_{L}^{D}$.

Because of the $T$-dependent renormalization, the parameters in the
reduced graph (and in the effective Lagrangian that generates such
graph) necessarily depend on $T$. This dependence is determined
uniquely by the original theory.
\vspace*{0.1cm}
\section{A model calculation}
Let us consider a concrete example, the Gross-Neveu model in 1+1
dimensions described by the Lagrangian
\begin{equation}
{\cal L} = \bar{\psi}i\gamma\cdot\partial\psi
-\bar{\psi}(\sigma+i\pi\gamma_5)\psi
-{N\over 2g^2}(\sigma^2+\pi^2)\, ,
\end{equation}
where $\sigma$ and $\pi$ are the auxiliary scalar and pseudoscalar
boson fields respectively. We study this model in the limit
$N\to\infty$ with the coupling constant $g^2$ fixed. By Fourier
transforming the fields
\begin{equation}
\psi(\tau,{x})={\sqrt{T}}\sum_{n=-\infty}^\infty
\psi_n({x})\/e^{i\omega_n\tau}\, , \quad
\begin{array}{c}
\sigma(\tau,{x}) \\ \pi(\tau,{x}) \end{array}
=\sum_{l=-\infty}^\infty
\begin{array}{c}
\sigma_l({x}) \\ \pi_l({x}) \end{array}
\/e^{i\Omega_l\tau}\, ,
\end{equation}
where $\omega_n=(2n-1)\pi T$ and $\Omega_l=2l\pi T$,
we can rewrite the action in the following way
\begin{equation}
\int d{x}\, \Bigg\{\sum_{n,l=-\infty}^\infty
\bar{\psi}_n({x})\bigg[-\omega_n\gamma_0
-i{\gamma_1}{\partial_1}-\sigma_l({x})
+i\gamma_5\pi_l({x})\bigg]\psi_n({x})
-{N\over 2 g^2T}\bigg[\sigma_l^2({x})+\pi_l^2({x})\bigg]
\Bigg\}\, .
\label{drl}
\end{equation}
Since we are only interested in Green's functions with zero external
frequency, only terms with $l=0$ are relevant. We are left with a
one dimensional fermion theory of infinite species, each with
a chirally invariant masses $\omega_n$. The tree-level coupling
constant is $g^2 T$. We expect that DR occurs, if the static
correlations are reproduced by the action (\ref{drl}) with only
$n=\pm 1$, and the higher modes at most modify the tree-level
effective theory in a local way, i.e. make the parameters
$T$-dependent.

In the $1/N$ limit and high-$T$ regime, this model has only one
non-trivial irreducible graph: the bubble graph. This graph in the
high-$T$ phase for static external lines is \cite{bernd}
\begin{eqnarray}
i\Pi(p_1)&=&-i{NT}\sum_{n=-\infty}^\infty\mu^{2\epsilon}
\int {d^{1-2\epsilon}k_1\over(2\pi)^{1-2\epsilon}}\,{\rm Tr}
\Bigg\{\gamma_5 {i\over k\cdot\gamma}
\gamma_5{i\over(k+p)\cdot\gamma}\Bigg\}\\
&=&-{N\over 2\pi}
\Bigg[{1\over\epsilon}
+\ln\bigg({4\mu^2\, e^{\gamma_E}\over\pi T^2}\bigg)
-{p_1^2\over T^2}\sum_{n=1}^\infty
\frac{1}{(2n-1)^3\pi^2 + (2n-1){p_1^2}/{4T^2}}\Bigg]\, .
\label{unren}
\end{eqnarray}
Let us renormalize the theory in a $T$-independent way. For instance,
the $\overline{\rm MS}$ scheme yields the renormalized coupling
${2\pi}/{g^2(\mu)}=\ln({4 \mu^2e^{\gamma_E}}/{\pi T^2_c})$
where $T_c\equiv{\Lambda_{\overline{MS}}e^{\gamma_E}}/{\pi}$.
Since we expect that the large distance behavior of the spatial
correlations is determined by the lowest singularities, i.e.
$p_1=\pm 2\pi T$, we analytically continue $p^2_1$ into Minkowski
space, and define the reduced momentum $q^2_1=-p_1^2 - 4\pi^2 T^2$.
Then Eq.~(\ref{unren}) becomes
\begin{eqnarray}
i\Pi(q^2_1)&=&-\frac{N}{\pi}\ln(\frac{T_c}{T})
-{2N\over \pi}{4\pi^2 T^2-q_1^2\over q_1^2}
\bigg[1+\sum_{n=1}^\infty
\frac{q_1^2}{(2n+1)[16\pi^2 T^2(n^2+n)-q_1^2]}\bigg]\, \nonumber \\
&=&-\frac{N}{\pi}\ln(\frac{T_c}{T})
-{2N\over \pi}{4\pi^2 T^2\over q_1^2}
\bigg[1+{\cal O}(\frac{q_1^2}{T^2})\bigg]\, .
\end{eqnarray}
This result illustrates what has been said in the previous section.
The main momentum dependence $-(2N/\pi)(4\pi^2 T^2/q_1^2)$ is given
by the lightest modes. The other modes give contributions that are
either suppressed by powers of ${q_1^2}/{T^2}$, or momentum
independent, $-{N}\ln({T_c}/{T})/\pi$. But this local piece can be
eliminated by a judicious choice of renormalization scale $\mu$,
e.g. $4\mu^2 ={\pi T^2}e^{-\gamma_E}$. Consequently, the coupling
that appears in any diagram of the original theory will effectively
depend on $T$ through:
\begin{equation}
g^2 \left. (\mu) \right|_{4\mu^2 e^{\gamma_E}={\pi T^2}}
= \frac{\pi}{\ln({T}/{T_c})}\equiv g^2(T) \, .
\label{gt}
\end{equation}
Obviously, the same $T$-dependence has to be inherited by the reduced
theory. One can explicitly check that this graph can be generated from
the reduced action, Eq.~(\ref{drl}) with $l=0$ and $n=\pm 1$, with
that $g^2$ replaced by $g^2(T)$ given by Eq.~(\ref{gt}).

In the original theory solving the equation $N/g^2-i\Pi=0$ yields
the screening mass
\begin{equation}
\tilde{m}={2\pi T}\bigg[1-{1\over\pi}g^2(T)
+{\cal O}(g^4(T))\bigg]\, .
\label{ms2}
\end{equation}
This result is reproduced by the reduced theory to leading order in
$g^2(T)$, and explains the physical reason why a partial DR works for
this theory. The screening state is a bound state of a quark and an
antiquark of mass $\pi T$, with a binding energy $\approx 2g^2(T)T$. The
binding energy in units of the quark mass decreases logarithmically,
so we expect that the screening mass can be solved from an appropriate
Schr\"{o}dinger equation in the high-$T$ limit. The point here is that
there exists a simplified physical picture, similar to the non-relativistic
reduction assumed by several authors~\cite{koch,hansson,schramm}.
In an asymptotically free theory, there is one more scale appearing
other than $T$. This new scale, ${T}/{\ln T}$, makes the partial DR
possible. On the contrary, in theories with a finite ultraviolet
fix-point, as we have explicitly verified in the 2+1 Gross-Neveu
model \cite{long94}, not even this partial dimensional reduction occurs,
due to the lack of the scale hierarchy.
\section{Conclusions and outlook}
We generalized the usual DR concept to the case of operators made by
fermion fields, where there is no obvious scale separation. Then, the
idea is illustrated in the Gross-Neveu model by solving the screening
mass in the high-$T$ limit. The experience gained in the model study
suggests that the complete DR (with corrections suppressed by powers of
$T$) in case of operators made by fermion fields is almost impossible,
since the overall scale is set by $T$. However, when the theory is
asymptotically free a new scale, $g^2(T)T$, becomes available, which
makes the partial DR possible.

Although the partial DR is the most likely scenario for operators
involving fermion fields in asymptotically free theories, the partial
reduction would still be sufficient
to guarantee a simplified physical picture, the non-relativistic
type of reduction, and thus is useful to interpret lattice data at
high-$T$ \cite{koch,hansson,schramm}. We are in the process of
explicitly examining whether a similar partial DR is going to occur
for currents made by fermions in QED and QCD.

We thank Profs.~H.~M.~Fried and B.~M\" uller for organizing a very
enjoyable workshop. This work is supported in part by the US
Department of Energy.

\vspace*{0.1cm}

\end{document}